\documentclass[twocolumn,preprintnumbers,amsmath,amssymb]{revtex4}

\usepackage{graphicx}
\usepackage{dcolumn}
\usepackage{bm}

\begin{document}


\title{Magnetization and spin dynamics of a Cr-based magnetic cluster: Cr$_{7}$Ni}

\author{A. Bianchi$^1$, S. Carretta$^1$, P. Santini$^1$, G. Amoretti$^1$, Y. Furukawa$^2$, K. Kiuchi$^2$,
Y. Ajiro$^3$, Y. Narumi$^4$, K. Kindo$^4$, J. Lago$^5$, E. Micotti$^5$, P. Arosio$^6$, A. Lascialfari$^{5,6}$, F. Borsa$^5$, G. Timco$^7$ and R. E. P. Winpenny$^7$}

\address{$^1$Dipartimento di Fisica, Universit\`a di Parma, Viale Usberti 7/A, I-43100 Parma, Italy; and S$^3$-CNR-INFM, I-41100 Modena, Italy}
\address{$^2$Division of Physics, Hokkaido University, Sapporo 060-0810, Japan}
\address{$^3$Department of Chemistry, Kyoto University, Kyoto 606-8502, Japan; and CREST Japan Science and Technology (JST)}
\address{$^4$The Institute for Solid State Physics, The University of Tokyo, Kashiwa, Chiba 277-8581, Japan}
\address{$^5$Dipartimento di Fisica \lq\lq A. Volta\rq\rq, Universit\`a di Pavia; and CNR-INFM, Via Bassi 6, I-27100 Pavia, Italy}
\address{$^6$Istituto di Fisologia Generale e Chimica Biologica, Universit\`a di Milano, I-20134 Milano, Italy; CNR-INFM, I-27100 Pavia, Italy; and S$^3$-CNR-INFM, I-41100 Modena, Italy}
\address{$^7$School of Chemistry, University of Manchester, Oxford Road, Manchester, M13 9PL United Kingdom}

\date{\today}

\begin{abstract}
We study the magnetization and the spin dynamics of the Cr$_7$Ni ring-shaped magnetic cluster. Measurements of the magnetization at high pulsed fields and low temperature are compared to calculations and show that the spin Hamiltonian approach provides a good description of Cr$_7$Ni magnetic molecule. In addition, the phonon-induced relaxation dynamics of molecular observables has been investigated. By assuming the spin-phonon coupling to take place through the modulation of the local crystal fields, it is possible to evaluate the decay of fluctuations of two generic molecular observables. The nuclear spin-lattice relaxation rate $1/T_1$ directly probes such fluctuations, and allows to determine the magnetoelastic coupling strength.
\end{abstract}


\maketitle

\section{Introduction}\label{Intro}

The great efforts devoted to the synthesis and investigation of nanosize magnetic molecules are motivated both by interests in fundamental physics and by the envisaged technological applications. For instance, some of these systems have shown phenomena such as quantum tunneling of magnetization between quasi-degenerate levels, slow relaxation at low $T$, and revealed to be promising for high density information storage and quantum computing \cite{Gatteschi,Sessoli,Leuenberger}.
The magnetic core of molecular magnets is constituted by transition metal ions sorrounded by an organic shell which prevents intramolecular magnetic interactions. As a result, the microscopic properties of these nanoscale clusters can be investigated by means of bulk samples. 
Among these systems, there are homonuclear antiferromagnetic (AF) ring-shaped molecules formed by $n$ transition metal ions in an almost planar ring. In particular, in even membered rings the dominant AF exchange interactions lead to a singlet $S_T=0$ ground state and the energy spectrum is characterized by rotational bands, with the lowest-lying levels approximately following the so-called Land\'e's rule \cite{Gatteschi}.
In this paper we study the magnetization and the phonon-induced relaxation in the heterometallic ring Cr$_7$Ni. This compound derives from the even membered AF ring Cr$_8$ and thus provides a opportunity of a deeper insight in the role of topology in the static and dynamical quantum properties of magnetic wheels \cite{CarrettaTopol,Furukawa}.
Cr$_7$Ni compound is obtained by the chemical substitution of a Cr$^{3+}$ ion with a Ni$^{2+}$ ion in the structure of the Cr$_8$ ring. This leads to a new molecular system formed by an odd number of unpaired electrons with dominant AF nearest neighbour interactions as inferred by susceptibility measurements \cite{Larsen}. The resulting $S_T=1/2$ ground state has been shown to be suitable to encode a qubit \cite{Troiani}.

\section{Magnetization}
\begin{figure}
	\centering
		\includegraphics[width=0.45\textwidth]{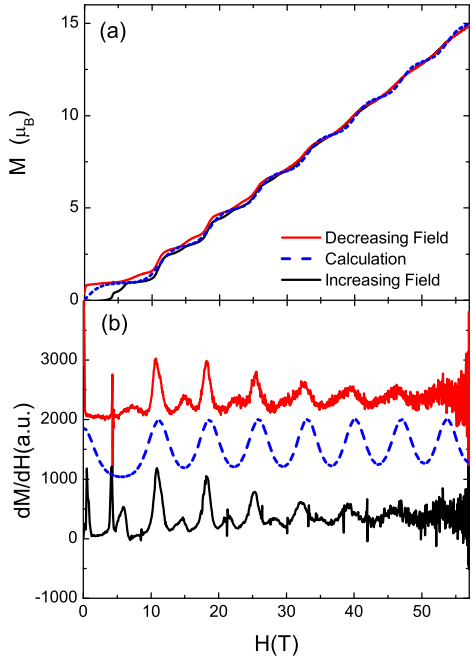}
	\caption{(Color online) Magnetization curve of Cr$_7$Ni (top) and derivative $dM/dH$ (bottom) at T=1.3K. Red (dark gray) and black lines represent the down and up experimental magnetic field processes respectively. The dashed blue lines represent the theoretical calculation with the following parameters: $J_{Cr-Cr}$=16.9K, $J_{Cr-Ni}$=19.6K, $d_{Cr}$=-0.3K, $d_{Ni}$=-4K, $g_{Cr}$=1.98, $g_{Ni}$=2.2.}
	\label{fig:Cr7NiM-article}
\end{figure}

The magnetic molecule has been theoretically analyzed within a spin Hamiltonian approach, with the Hamiltonian given by:
\begin{eqnarray}
H=\sum_{i>j}{J_{ij}\textbf{s}(i)\cdot\textbf{s}(j)}+\sum_i{d_i\left[s^{2}_{z}(i)-s_i(s_i+1)/3\right]}\nonumber\\
+\sum_{i>j}\textbf{s}(i)\cdot\textbf{D}_{ij}\cdot\textbf{s}(j)-\mu_B\sum_i{g_{i}\textbf{H}\cdot\textbf{s}(i)},
\end{eqnarray}
where $\textbf{s}(i)$ is the spin operator of the the $i$th ion in the molecule ($s(i)$=3/2 for Cr$^{3+}$ ions, and $s(i)$=1 for the Ni$^{2+}$ ion). The first term of the above equation is the dominant nearest neighbour Heisenberg exchange interaction. The second and the third terms describe the uniaxial local crystal fields and anisotropic intracluster spin-spin interactions respectively (with the $z$ axis assumed perpendicular to the plane of the ring). The last term represents the Zeeman coupling with an external field $\textbf{H}$. The parameters of the above Hamiltonian were determined by inelastic neutron scattering (INS) experiments \cite{Caciuffo,CarrettaPRL}. In order to corroborate the microscopic description of the Cr$_7$Ni from INS data, a detailed study of high field magnetization is very powerful. In fact, with high pulsed fields up to almost 60T, spin multiplets not accessible to a standard INS experiment can be explored. In Fig.\ref{fig:Cr7NiM-article}a the magnetization curve as a function of the magnetic field $\textbf{H}$ is reported. A clear staircase structure with plateaus at $\approx 1\mu_B$, $\approx 3\mu_B$ and odd multiples of $\mu_B$ reflects the change in the ground state due to the external field at the level anticrossing fields $H_n$. An hysteresis of the measured magnetization curves has been observed. The effect arises from the non-equilibrium condition due to the high pulsed magnetic field with a few millisecond duration \cite{Furukawa} and has been discussed in terms of phonon bottle-neck effects and magnetic Foehn effects \cite{Nakano}. There is a very good agreement between the measured and calculated magnetization curves. This is clearly visible in Fig.\ref{fig:Cr7NiM-article}b where the positions of the main peaks of the calculated and measured $dM/dH$ matches correctly. The smaller peaks in the experimental $dM/dH$ are due to level anticrossings between excited energy levels. The effects are caused by the non-equilibrium exeperimental conditions and are not included in equilibrium calculations reported in Fig.\ref{fig:Cr7NiM-article} \cite{Furukawa}. These results confirm that the microscopic picture derived from INS experiments \cite{Caciuffo,CarrettaPRL} perfectly holds even at very high applied magnetic fields.

\section{Spin dynamics}
A major obstacle to the proposed technological applications of magnetic molecules is constituted by phonon-induced relaxation. In fact, molecular observables, e.g. the magnetization, are deeply affected by the interaction of the spins with other degrees of freedom such as phonons \cite{SantiniNMR}. 
Here we investigate the molecular spin-spin correlations through an approach based on a density matrix theory \cite{SantiniNMR}. The irreversible evolution of the density matrix $\hat{\rho}(t)$ can be determined through the secular approximation and focusing on time scales detectable by low-frequency techiques such as NMR. Within this theoretical framework, a general expression for the quasi-elastic part of the Fourier transform of cross correlation functions is given by \cite{SantiniNMR,Bertaina}:
\begin{eqnarray}\label{eq:S}
	S_{AB}(\omega)=\sum_{m,n=1,N}p^{(eq)}_{m}(B_{mm}-\left\langle \hat{B}\right\rangle_{eq})\nonumber\\ \times(A_{nn}-\left\langle\hat{A}\right\rangle_{eq})
	Re\left\{\left(\frac{1}{i\omega-W}\right)_{nm}\right\},
\end{eqnarray}
where $N$ is the dimension of the Hilbert spin space of the molecule, $p^{(eq)}_{m}$ is the equilibrium population of the $m$th level and $B_{mm}=\left\langle m\right|\hat{B}\left|m\right\rangle$, $\left|m\right\rangle$ being the $m$th eigenstate of the spin Hamiltonian, while $W$ is the so called rate-matrix. The $mn$ element $W_{mn}$ of $W$ represents the probability per unit time of a transition between eigenstates $\left|m\right\rangle$ and $\left|n\right\rangle$ induced by the interaction of the spins with phonons. By assuming that spin-bath interaction takes place mostly through modulation of local crystal fields, the rate matrix can be calculated on the basis of the eigenstates of molecular spin Hamiltonian by first-order perturbation theory. With the choice of a spherical magnetoelastic (ME) coupling \cite{CarrettaNi10} the transition rates $W_{mn}$ are given by:
\begin{eqnarray}\label{eq:W}
	W_{mn}=\gamma\pi^{2}\Delta^{3}_{mn}n(\Delta_{mn})\sum_{\stackrel{i,j=1,N}{q_1,q_2=x,y,z}}\left\langle m\right|O_{q_1,q_2}(\textbf{s}_i)\left|n\right\rangle
\nonumber\\ 
\times\left\langle n\right|O_{q_1,q_2}(\textbf{s}_j)\left|m\right\rangle,
\end{eqnarray}
with $n(x)=(e^{\beta\hbar x}-1)^{-1}$, $\Delta_{mn}=(E_m-E_n)/\hbar$ the gap between the eigenstates $\left|m\right\rangle$ and $\left|n\right\rangle$ of the molecule.
In the last equation $O_{q_1,q_2}(\textbf{s}_i)=(s_{q1,i}s_{q2,i}+s_{q2,i}s_{q1,i})/2$ are quadrupolar operators \cite{CarrettaNi10}. Finally, $\gamma$ represents the spin-phonon coupling strength, which can be determined by comparing the theoretical results with experimental data. In fact, the nuclear spin-lattice relaxation rate $1/T_1$ probes the fluctuations of molecular observables, thus giving information on the relaxation dynamics \cite{SantiniNMR}. Exploiting the Moriya formula \cite{Moriya}, the proton NMR $1/T_1$ can be evaluated in absolute units using as inputs the positions of the Cr and Ni ions and of the hydrogens of the molecule:
\begin{equation}{\label{eq:T1}}
\frac{1}{T_1}=\sum_{\stackrel{i,j=1,N}{q,q'=x,y,z}}\alpha^{qq'}_{ij}
\left(S_{s^{q}_{i},s^{q'}_{j}}(\omega_L)+S_{s^{q}_{i},s^{q'}_{j}}(-\omega_L)\right),	
\end{equation}
where the $S_{s^{q}_{i},s^{q'}_{j}}(\omega_L)$ are the Fourier transforms of the cross correlation functions from Eq. (\ref{eq:S}) calculated at the Larmor angular frequency $\omega_L$, while the $\alpha^{qq'}_{ij}$ are geometric coefficients of the hyperfine dipolar interaction between magnetic ions and protons probed by NMR. 
\begin{figure}
	\centering
		\includegraphics[width=0.45\textwidth]{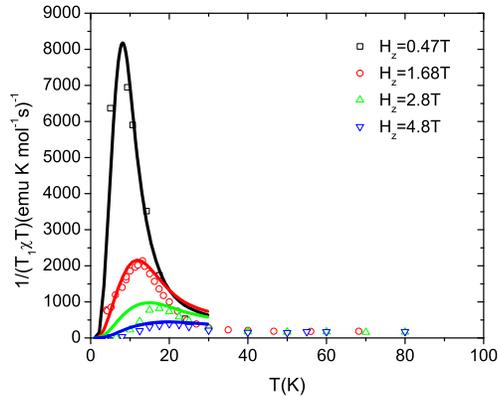}
	\caption{(Color online) Experimental data (scatters) and calculations (lines) of reduced proton NMR $1/(T_1\chi T)$ for different values of the applied field along z (parallel to the ring axis).}
	\label{fig:Cr7Ni-T1-Hz}
\end{figure}
The occurence of a peak in the proton NMR $1/T_1$ has been clearly explained in homonuclear ring-shaped molecules with small anisotropy such as Cr$_8$ \cite{SantiniNMR}. In fact, in this case $1/T_1\propto S_{S_z,S_z}(\omega_L)$, where $S_{S_z,S_z}(\omega,\textbf{H},T)$ is the Fourier transform of the autocorrelation function of $M$ \cite{SantiniNMR}:
    $S_{S_{z},S_{z}}(\omega,T,\textbf{H})=\sum_{i=1,N}A(\lambda_{i},T,\textbf{H})\lambda_{i}(T,\textbf{H})/
    [\lambda_{i}(T,\textbf{H})^{2}+\omega^{2}]$.
This equation shows that the spectrum of fluctuations of $M$ is given by a sum of $N$ Lorentzians, each with characteristic frequency $\lambda_i$, given by the eigenvalues of $-W$. For a wide range of $\textbf{H}$ and $T$ in these systems only a single relaxation frequency $\lambda_0$ significantly contribute to $S_{S_{z},S_{z}}(\omega,T,\textbf{H})$. As a result, if the dominant frequency $\lambda_0$ intersects the Larmor angular frequency, i.e. when $\lambda_0(T_0)=\omega_L$, at the temperature $T_0$ the proton NMR $1/T_1$ shows a sharp peak \cite{SantiniNMR,Baek}. Being an heterometallic ring, this explanation does not hold for Cr$_7$Ni and Eq.(\ref{eq:T1}) has to be used. Nevertheless, our calculations show that a peak in the reduced $1/(T_{1}\chi T)$ occurs in agreement with experimental data (see Fig.\ref{fig:Cr7Ni-T1-Hz}). By fitting the observed peak position we have obtained $\gamma=0.8\times10^{-7}$THz$^{-2}$.
\section{Conclusions}
A magnetization study of the heteronuclear antiferromagnetic ring-shaped nanomagnet Cr$_7$Ni has been performed. A clear step-wise increase of magnetization with increasing field is observed. The very good agreement of high field magnetization measurements up to almost 60T with calculation shows the spin Hamiltonian approach to be suitable even at very high fields.
The relaxation dynamics of the compound has been investigated by the proton nuclear-spin relaxation rate $1/T_1$. Our calculations are in very good quantitative agreement with experimental data. 




\begin{references}



\bibitem{Gatteschi}D. Gatteschi, R. Sessoli, and J. Villain, \textit{Molecular Nanomagnets}, Oxford University Press, Oxford (2006).
\bibitem{Sessoli}R. Sessoli, D. Gatteschi, A. Caneschi, and M. A. Novak, Nature (London) \textbf{365}, 141 (1993).
\bibitem{Leuenberger}M. N. Leuenberger and D. Loss, Nature (London) \textbf{410}, 789 (2001).
\bibitem{Larsen} F. K. Larsen, E. J. L. McInnes, H. El Mkami, J. Overgaard, S. Piligkos, G. Rajaraman, E. Rentschler, A. A. Smith, G. M. Smith, V. Boote, M. Jennings, G. A. Timco, and R. E. P. Winpenny, Ang. Chemie \textbf{42}, 101 (2003).
\bibitem{Troiani}F. Troiani, A. Ghirri, M. Affronte, S. Carretta, P. Santini, G. Amoretti, S. Piligkos, G. Timco, and R. E. P. Winpenny, Phys. Rev. Lett. \textbf{94}, 207208 (2005).
\bibitem{CarrettaTopol} S. Carretta, P. Santini, G. Amoretti, M. Affronte, A. Ghirri, I. Sheikin, S. Piligkos, G. Timco, and R. E. P. Winpenny, Phys. Rev. B \textbf{72}, 060403 (2005).
\bibitem{Furukawa} Y. Furukawa et al., to be published.
\bibitem{Caciuffo}R. Caciuffo, T. Guidi, G. Amoretti, S. Carretta, E. Liviotti, P. Santini, C. Mondelli, G. Timco, C. A. Muryn, and R. E. P. Winpenny, Phys. Rev. B \textbf{71}, 174407 (2005).
\bibitem{CarrettaPRL} S. Carretta, P. Santini, G. Amoretti, T. Guidi, J. R. D. Copley, Y. Qiu, R. Caciuffo, G. Timco, and R. E. P. Winpenny, Phys. Rev. Lett. \textbf{98}, 167401 (2007).
\bibitem{SantiniNMR}P. Santini, S. Carretta, E. Liviotti, G. Amoretti, P. Carretta, M.
Filibian, A. Lascialfari, and E. Micotti, Phys. Rev. Lett. \textbf{94}, 077203 (2005).
\bibitem{Baek} S.-H. Baek, M. Luban, A. Lascialfari, E. Micotti, Y. Furukawa, F. Borsa, J. van Slageren, 
and A. Cornia, Phys. Rev. B \textbf{70}, 134434 (2004).
\bibitem{Nakano}H. Nakano and S. Miyashita, J. Phys. Soc. Jpn. \textbf{71}, 2580 (2002).
\bibitem{Bertaina} S. Bertaina, B. Barbara, R. Giraud, B. Z. Malkin, M. V. Vanuynin, A. I. Pominov, A. L. Stolov, and A. M. Tkachuk, Phys. Rev. B \textbf{74}, 184421 (2006).
\bibitem{CarrettaNi10} S. Carretta, P. Santini, G. Amoretti, M. Affronte, A. Candini, A. Ghirri, I. S. Tidmarsh,
R. H. Laye, R. Shaw, and E. J. L. McInnes, Phys. Rev. Lett. \textbf{97}, 207201 (2006).
\bibitem{Moriya} T. Moriya, Progr. Theor. Phys. \textbf{16}, 23 (1956). 

\end{references}
\end{document}